\documentclass[todo]{myarticle}
\DeclareMathOperator{\sgn}{sgn}

\usepackage{authblk}

\title{Gauge Symmetry Breaking Lattice Regularizations\\and their Continuum Limit}

\author[1]{Thorsten Lang\thanks{\texttt{thorsten.lang@fau.de}}}
\affil[1]{Institute for Quantum Gravity, FAU Erlangen--Nürnberg, Staudtstraße 7/B2, 91058 Erlangen, Germany}
\author[2]{Susanne Schander\thanks{\texttt{sschander@perimeterinstitute.ca}}}
\affil[2]{Perimeter Institute, 31 Caroline St N, Waterloo, ON N2L 2Y5, Canada}
\date{\today}

\begin{document}
\maketitle

\begin{abstract}
    Lattice regularizations are pivotal in the non--perturbative quantization of gauge field theories.
    Wilson's proposal to employ group-valued link fields simplifies the regularization of gauge fields in principal fiber bundles, preserving gauge symmetry within the discretized lattice theory.
    Maintaining gauge symmetry is desirable as its violation can introduce unwanted degrees of freedom.
    However, not all theories with gauge symmetries admit gauge--invariant lattice regularizations, as observed in general relativity where the diffeomorphism group serves as the gauge symmetry.
    In such cases, gauge symmetry--breaking regularizations become necessary.
    In this paper, we argue that a broken lattice gauge symmetry is acceptable as long as gauge symmetry is restored in the continuum limit.
    We propose a method to construct the continuum limit for a class of lattice--regularized Hamiltonian field theories, where the regularization breaks the Lie algebra of first--class constraints.
    Additionally, we offer an approach to represent the exact gauge group on the Hilbert space of the continuum theory. 
    The considered class of theories is limited to those with first--class constraints linear in momenta, excluding the entire gauge group of general relativity but encompassing its subgroup of spatial diffeomorphisms.
    We discuss potential techniques for extending this quantization to the full gauge group.
\end{abstract}
\tableofcontents*

\section{Introduction}
Establishing rigorous definitions of quantum field theories in the continuum presents a formi\-dable challenge.
It arises from the presence of infinities within naively constructed interaction terms and is rooted in the generally ill--defined process of multiplying distributions. 
Consequently, it is not feasible to directly formulate quantum field theories in the continuum. 
Instead, a common approach is to first establish regularized versions of the theory as an intermediate step. 
These regularized theories encompass a distinct set of parameters to govern the regularization process, supplementing the inherent free parameters of the theory. 
Subsequently, the continuum theory emerges as a limit reached through a carefully tuned trajectory in the theory space. Along this path, divergent terms are compensated by appropriately chosen counterterms, ensuring the existence of the limit \parencite{Montvay:1994cy}.

Facilitating the existence of the limit is greatly enhanced by preserving key attributes of the continuum theory within the regularized theories.
Although, in principle, certain features may and sometimes have to manifest in the limit despite their absence at finite (lattice) orders, it proves significantly more straightforward to seamlessly inherit these features from the regularized theories. 

An illustrative instance demonstrating the challenge of preserving a feature after the regularization process pertains to the problem of fermion doubling in lattice field theory. 
In fact, due to a theorem by \textcite{Nielsen:1981hk}, lattice field theories cannot accommodate chiral fermions --- an essential component of the standard model in particle physics. 
To circumvent this issue, a widely adopted approach involves acknowledging the existence of the spurious doubler degrees of freedom on the lattice while progressively suppressing them as the continuum limit is approached. 

A similar phenomenon arises when a gauge symmetry is broken on a lattice. 
In the Hamiltonian formalism, the existence of gauge symmetry is characterized by the presence of a system of first--class constraints. 
The lattice regularization process breaks the first--class property, leading to a situation where the constraints are no longer conserved during time evolution. 
Consequently, physical states eventually acquire unphysical degrees of freedom. 
While the origin of these unphysical degrees of freedom differs from the fermion doubling phenomenon, the effect is similar and poses comparable challenges. 

Unlike the issue of fermion doubling, breaking gauge symmetries on a lattice can often be circumvented. 
A method introduced by \textcite{Wilson:1974sk} allows to regularize a gauge field in terms of group--valued link variables on a lattice, representing the holonomies of the gauge field. 
From these link variables, one can construct gauge--invariant entities around closed loops known as Wilson loops, which are then utilized to obtain regularized and invariant representations of gauge theory actions.
A notable example of this approach is the Wilson action, which serves as a gauge--invariant lattice regularization of the Yang--Mills action. 

We would like to emphasize that this lattice regularization scheme only applies to gauge fields defined in terms of connections on principal $G$--bundles and when gauge transformations align with principal $G$--automorphisms. 
While this framework is well--suited for many theories of interest, including Yang--Mills theory and the standard model, it is inapplicable in the case of general relativity.
In principle, it's feasible to reformulate the theory, using the frame bundle, in the language of principal $GL(n)$--bundles and connections thereon. 
However, a crucial distinction arises in the group of gauge transformations, given by the group of spacetime diffeomorphisms. 
In general, these diffeomorphisms do not correspond to principal $GL(n)$--automorphisms on the frame bundle as they don't preserve the base manifold \parencite{Isham:1999rh}.
Consequently this hinders the direct implementation of Wilson's techniques for obtaining a lattice regularization of general relativity while maintaining gauge invariance.

The enduring prevalence of the Wilson action (and its extensions) as the primary choice for a lattice formulation of Yang--Mills theory is no mere coincidence. 
The inherent gauge invariance of this action serves to effectively prevent the emergence of unwanted physical degrees of freedom and greatly contributes to the theory's renormalizability. 
However, as appealing as the preservation of gauge invariance may be, there exists room for exploration into alternative regularization schemes that do break gauge invariance. 
Indeed, the literature contains various attempts in this direction \parencite[see, for instance,][]{Rivasseau-YM4}. 
Similar to the previously mentioned case of fermion doubling, alternative approaches may initially introduce unphysical degrees of freedom at the regularization level, which must subsequently be carefully controlled and suppressed. 
While the introduction of these additional technical challenges should not be underestimated, it does not diminish the general viability of such alternative methods. 

In fact, our need for an alternative approach arises from the inability to construct a gauge--invariant lattice regularization for general relativity, prompting us to explore a different path. 
In our previous work \parencite{lattice}, we introduced a lattice regularization for the Hamiltonian formulation of classical general relativity employing spatial metric variables. 
See also \textcite{Regge,spinfoams,CDT} for other approaches of discretizing gravity.
The natural progression toward a quantum gravity theory first necessitates the quantization of the regularized expression, followed by the challenging task of taking the continuum limit. 
This paper aims to provide an overview of our proposed approach for achieving this limit, specifically addressing the representation of the diffeomorphism group, derived from the diffeomorphism constraints, on the resulting Hilbert space.
To enhance the readability of this outline paper, we opt not to delve directly into the intricate details arising from our prior work \parencite{lattice}. 
Instead, we explore a simplified scenario where the metric tensor is replaced by a scalar field, and a general set of constraints is considered. 
While these restrictions may not directly apply to many intriguing theories, their extension to general relativity should become evident.

A significant limitation of this method lies in the necessity for the regularized constraints to be linear in the momenta.
Although this requirement might initially appear stringent, it finds compliance within a wide range of symmetry generators in physics.
This criterion ensures that the generated symmetries are defined by their action on the configuration variables, with their action on the momenta already prescribed.
In the context of gravity, this condition holds true for the diffeomorphism constraints that generate spatial diffeomorphisms, but not for the Hamiltonian constraint. 

Our current approach addresses only a portion of the challenge at hand. 
While the Hamiltonian constraint can be quantized on the lattice utilizing the technique introduced in \textcite{cholesky}, bridging the gap to its continuum limit necessitates a distinct undertaking. 
While the regularized algebra of spatial diffeomorphism constraints encounters gauge--breaking anomalies solely proportional to the lattice spacing, the complete regularized Dirac algebra incorporates terms proportional to $\hbar$.
These would persist in a naive limit of an infinitesimal lattice spacing.
Consequently, we anticipate the need for the application of renormalization group methods (e.g., \textcite{Lang:2017beo,Thiemann-Renormalization}, see also \textcite{Asante:2022dnj,Ambjorn:2020rcn,Saueressig:2023irs} for overviews of renormalization schemes in path integral approaches to quantum gravity\footnote{A more detailed overview of renormalization in quantum gravity will be provided in \textcite{gravity-continuum} along with an extensive number of references.}), to produce suitable counterterms.
Nevertheless, we envision our proposal as providing an ideal foundation for such analyses. 
As astute readers may discern, our method seemingly holds the potential to produce non--trivial continuum Hilbert spaces with corresponding representations of the gauge group. 

Lastly, it is worth emphasizing that this paper outlines the general framework of our method without delving into the technical for every statement. 
We have chosen to reserve these details for an upcoming paper \parencite{gravity-continuum}, where we will provide a more concrete and comprehensive treatment. 

With this, let us come to the structure of this paper:
In \cref{sec:General Setup}, we present a comprehensive review of the regularization and the quantization of the Hamiltonian field theory under examination.
This deliberately aligns with the findings of \textcite{lattice}.
In contrast to the findings of \textcite{cholesky}, we use a standard Schrödinger representation for the purpose of illustration.
Subsequently, \Cref{sec:The Continuum Limit} examines methods for obtaining non--trival continuum limits of the discretized and quantized theories.
Finally, in \cref{sec:Conclusion}, we offer a discussion of our results, summarizing our key findings.

\section{General Setup}
\label{sec:General Setup}

\subsection{Classical Theory}
For simplicity, let us consider a classical one--dimensional Hamiltonian field theory of a scalar field $\phi(x)$ and its canonically conjugate momentum $\pi(x)$ on a torus $\mathbb T = [0,1]$, where $0$ and $1$ are identified.
We assume that there exists an algebra of constraints $D[f]$ with
\begin{equation}
    D[f] = \int_{\mathbb T} \mathcal D(\phi(x), \partial \phi(x), \pi(x), \partial \pi(x)) f(x) \d x 
\end{equation}
such that the first class relations
\begin{equation}
    \poisson{D[f]}{D[g]} = D[F(f,\partial f,g,\partial g)] \label{eq:classical-continuum-algebra}
\end{equation}
hold for some $F$ that does not depend on the fields $\phi$ or $\pi$.

The aim is to derive a lattice discretized version of this theory on a regular lattice with lattice spacing $\eta$.
In order to achieve, this, we proceed in two steps.
First, we replace all occurring spatial derivatives in $D[f]$, such as $\partial f(x)$, by finite differences
\begin{equation}
    \Delta^\eta f(x) \coloneq \frac{f(x+\eta)-f(x)}\eta,
\end{equation}
and obtain approximate expressions $D_\eta[f]$.
In a second step, we evaluate the resulting expression on a restricted set of phase space functions, parameterized only by a finite number of degrees of freedom.
On a lattice with lattice spacing $\eta_n = 2^{-n}$, we define a piecewise constant version of the field given by
\begin{equation}
    \phi_n(x) \coloneq \sum_{k=1}^{N_n} \phi_{nk} \, \chi_{X_k}(x),
\end{equation}
where $N_n = 2^n$, $X_k = [(k-1) 2^{-n}, k 2^{-n}]$ with $k = 1, \dots, 2^n$, and $\chi_{X_k}$ is the characteristic function on $X_k$.
Similarly, we define piecewise constant versions of $\pi$ and $f$.

Strictly speaking, these functions do not really arise as a restriction of the original phase space, because the latter consists of differentiable functions.
However, our choice of the phase space is a useful way of understanding the process.
Of course, it is possible to use piecewise polynomial, everywhere differentiable functions instead.
This would constitute a higher--order approximation and a true restriction of phase space.
It would even allow us to skip the first step of replacing derivatives by finite differences.
The zeroth order approximation using piecewise constant functions, however, turns out to be most convenient and sufficient in order to obtain a lattice approximation of the continuum theory.

By evaluating the approximate constraint $D_{\eta_n}[f]$ on the piecewise constant functions $\phi_n$, $\pi_n$ and $f_n$, we obtain a version of the constraints that only depends on the finitely many degrees of freedom $(\phi_{nk})_k$ and $(\pi_{nk})_k$:
\begin{align}
    \MoveEqLeft D_{\eta_n}[f_n]
         = \int_{\mathbb T} \mathcal D(\phi_n(x), \Delta^{\eta_n} \phi_n(x), \pi_n(x), \Delta^{\eta_n} \pi_n(x)) f_n(x) \d x \\
        &= \sum_{k=1}^{N_n} \int_{X_k} \mathcal D(\phi_n(x), \Delta^{\eta_n} \phi_n(x), \pi_n(x), \Delta^{\eta_n} \pi_n(x)) f_n(x) \d x \\
        &= \sum_{k=1}^{N_n} \mathcal D(\phi_{nk}, \Delta^{\eta_n} \phi_{nk}, \pi_{nk}, \Delta^{\eta_n} \pi_{nk}) f_{nk} \eta_n . \label{eq:constraint-lattice}
\end{align}
By abuse of notation, we defined
\begin{equation}
    \Delta^{\eta_n} \phi_{nk} \coloneq \frac{\phi_{n,k+1} - \phi_{n,k}}{\eta_n} .
\end{equation}
\Cref{eq:constraint-lattice} represents a well defined expression on a finite--dimensional phase space with canonical variables $(\phi_{nk})_k$ and $(\pi_{nk})_k$.
Since we choose the underlying manifold to be a torus $\mathbb T$, we will assume periodic boundary conditions such as $\phi_{n,N_n+1} = \phi_{n,1}$.
Moreover, we will restrict ourselves to the lattice spacings $\eta_n = 2^{-n}$ as introduced above in the remainder of the paper.
Moreover, we will assume that $D_n$ is sufficiently well behaved such that $D_n[f_n]\to D[f]$ as $n\to\infty$ if an appropriate sequence $f_n(x)\to f(x)$ is chosen.

The Poisson bracket on the phase space of piecewise constant functions must be chosen in order to be consistent with the continuum Poisson bracket
\begin{equation}
    \poisson{\phi[f]}{\pi[g]} = \int_{\mathbb T} f(x)\, g(x) \d x .
\end{equation}
By evaluating this expression on piecewise constant functions, we find \parencite[cf.][]{lattice}
\begin{equation}
    \poisson{\phi_{nk}}{\pi_{nk'}} = \eta_n^{-1} \delta_{kk'} . \label{eq:classical-lattice-ccr}
\end{equation}
This allows us to compute the Poisson brackets of the lattice constraints $D_n[f_n]$.
We will assume that they obey the relation
\begin{equation}
    \poisson{D_n[f_n]}{D_n[g_n]} = D_n[F_n(f_n,\Delta^n f_n,g_n,\Delta^n g_n)] + \eta_n G_n(f_n,\Delta^n f_n,g_n,\Delta^n g_n) \label{eq:classical-lattice-algebra}
\end{equation}
with functions $F_n$ such that $D_n[F_n]\to D[F]$, and $G_n$ is bounded as $n\to\infty$.
This assumption holds, for instance, in the case of the diffeomorphism constraints, as was shown in \textcite{lattice}.

Let us interpret this relation.
Due to our assumptions, the $\eta_n G_n$ term will approach zero as $n$ increases.
Thus, for sufficiently fine lattices, the resulting algebra is dominated by the $D_n$ terms.
As $D_n[f_n]\to D[f]$, the algebra more and more resembles the original continuum algebra \labelcref{eq:classical-continuum-algebra}.
The continuum constraint $D[f]$ generates gauge transformations by the evolution according to Hamilton's equations of motion
\begin{equation}
    \diff{\phi[g]}{s} = \poisson{\phi[g]}{D[f]}, \quad \diff{\pi[g]}{s} = \poisson{\pi[g]}{D[f]}.
\end{equation}
\Cref{eq:classical-lattice-algebra} implies that we can interpret the Hamiltonian evolution
\begin{equation}
    \diff{\phi_n[g_n]}{s} = \poisson{\phi_n[g_n]}{D_n[f_n]}, \quad \diff{\pi_n[g_n]}{s} = \poisson{\pi_n[g_n]}{D_n[f_n]} \label{eq:hamiltons-equations}
\end{equation}
of the lattice degrees of freedom with respect to $D_n[f_n]$ as lattice approximations of continuum gauge transformations.

It is important to contemplate for a moment the presence of the $G_n$ terms in the lattice algebra \labelcref{eq:classical-lattice-algebra}.
Strictly speaking, their existence makes the lattice constraints $D_n[f_n]$ into second class constraints, since their Poisson bracket ceases to be a linear combination of constraints.
This means that the evolution of a point in phase space with respect to an approximate gauge transformation will move that point out of the constraint hypersurface.
Thus, states will acquire unphysical degrees of freedom, even if they initially satisfied the constraints.
Moreover, the composition of two approximate gauge transformations can not again be written as an exact approximate gauge transformation with respect to a constraint $D_n[h_n]$ for some $h_n$ and thus, they will not form a group.

However, due to the smallness of $\eta_n G_n$, the composition of two approximate gauge transformations will again be very close to an approximate gauge transformation of the same class, at least for sufficiently short evolution periods.
Also, and again for short evolution periods, the condition $D_n[f_n]\to D[f]$ ensures that approximate gauge transformations will resemble actual continuum gauge transformations quite well as the lattice approaches the continuum.
In order to control the error bounds, one may have to let the lattice spacing $\eta_n$ approach zero more quickly than one needs to let the sequence $f_n$ of lattice test functions approach the continuum test function $f$, or in other words, in order to obtain a good approximation to a continuum gauge transformation, one needs to work on a sufficiently fine lattice.

For the remainder of the paper, we will assume that $D_n[f_n]$ depends only linearly on the momenta $\pi_{nk}$.
As a consequence, Hamilton's equation of motion \labelcref{eq:hamiltons-equations} for $\phi_n[g_n]$ is a first order system of differential equations that only depends on the configuration variables $\phi_{nk}$ and not on the momenta.
Thus, its solution only requires initial data for the configuration variables.

The map $\varphi_s^{D_n[f_n]} \colon \mathbb R^n\to\mathbb R^n$ that sends an initial configuration to its evolved version with regards to the evolution with respect to $D_n[f_n]$ and with evolution parameter $s$, is called the Hamiltonian flow.
It satisfies $\varphi_0^{D_n[f_n]} = \id$, is invertible with inverse given by $\cramped{\left(\varphi_s^{D_n[f_n]}\right)^{\mathrlap{-1}} = \varphi_{-s}^{D_n[f_n]}}$ and respects the composition rule $\varphi_s^{D_n[f_n]} \circ \varphi_t^{D_n[f_n]} = \varphi_{s+t}^{D_n[f_n]}$.
The Hamiltonian flow of $D_n[f_n]$ is a canonical transformation on phase space and represents the approximate gauge transformation generated by it.

\subsection{Quantum Theory}
We will now proceed with the quantization of the above--introduced classical lattice field theories, focusing on the Schrödinger representation as an illustrative case. 
Nonetheless, it is worth noting that alternative representations may prove advantageous depending on the circumstances \parencite[cf.]{cholesky}.
Within the Schrödinger representation, the Hilbert space associated with a theory featuring a lattice spacing $\eta_n$ is defined as $\mathcal H_n = L^2(\mathbb R^n)$.
To quantize the field degrees of freedom $\phi_{nk}$, we employ multiplication operators
\begin{equation}
    (\hat\phi_{nk}\psi_n)(\phi_{n1},\ldots,\phi_{n N^n}) = \phi_{nk}\psi_n(\phi_{n1},\ldots,\phi_{n N^n}),
\end{equation}
while the momentum degrees of freedom $\pi_{nk}$ act as differential operators according to
\begin{equation}
    (\hat\pi_{nk}\psi_n)(\phi_{n1},\ldots,\phi_{n N^n}) = -\im\eta_n^{-1}\diffp{\psi_n}{\phi_{nk}}(\phi_{n1},\ldots,\phi_{n N^n}) .
\end{equation}
We emphasize the factor $\eta_n^{-1}$, which ensures the correct quantization of the classical Poisson bracket relation \labelcref{eq:classical-lattice-ccr}
\begin{equation}
    \commutator{\hat\phi_{nk}}{\hat\pi_{nk'}} = \im\eta_n^{-1} \delta_{kk'} .
\end{equation}

We then proceed to define a representation of the approximate gauge transformations $\varphi_s^{D_n[f_n]}$ on the lattice Hilbert space $\mathcal H_n$ as follows\footnote{A similar construction has been used by \textcite{Thiemann-U1} for the quantization of Euclidean $U(1)$ gravity in a Narnhofer--Thirring type representation.}:
\begin{equation}
    \left(U\,\left(\varphi_s^{D_n[f_n]}\right)\psi_n\right)((\phi_{nk})_k) = \sqrt{\det\left(J_{\varphi_s^{D_n[f_n]}}((\phi_{nk})_k)\right)}\,\psi_n(\varphi_s^{D_n[f_n]}((\phi_{nk})_k)). \label{eq:quantum-gauge-transformation}
\end{equation}
For simplicity, let us denote this expression by $U_{D_n[f_n]}(s)$ from now on.
Here, $\cramped{J_{\varphi_s^{D_n[f_n]}}}$ is the Jacobian matrix of the Hamiltonian flow.

This representation inherits the properties of the Hamiltonian flow $\varphi_s^{D_n[f_n]}$:
Evidently, it evaluates to the identity transformation for $s=0$, and a simple application of the multivariable chain rule immediately yields
\begin{equation}
    U_{D_n[f_n]}(s)\,U_{D_n[f_n]}(t)=U_{D_n[f_n]}(s+t) ,
\end{equation}
which also implies that $(U_{D_n[f_n]}(s))^{-1} = U_{D_n[f_n]}(-s)$.
The square root in \cref{eq:quantum-gauge-transformation} ensures that $\norm{U_{D_n[f_n]}(s)} = 1$.
Taken together, we find that $U_{D_n[f_n]}(s)$ is a unitary operator.
Moreover, the continuity of $\varphi_s^{D_n[f_n]}$ in $s$ implies that $U_{D_n[f_n]}(s)$ is strongly continuous, as is evident from \cref{eq:quantum-gauge-transformation}.
Therefore, we have obtained a strongly continuous one--parameter group of transformations, which, according to Stone's theorem, also possesses a generator, which can be taken to be the representation of the lattice constraint $D_n[f_n]$ on the lattice Hilbert space:
\begin{equation}
    \hat D_n[f_n] \psi_n \coloneq -\im\diff{U_{D_n[f_n]}(s)}s[\mathrlap{s=0}]\; \psi_n .
\end{equation}

We emphasize that for this construction to work, it is essential that $D_n[f_n]$ depends only linearly on the momenta $\pi_{nk}$.
Otherwise, the Hamiltonian flow of the configuration variables would depend not only on the configuration variables, but also on the momenta as initial conditions.
Since the $\psi_n\in\mathcal H_n$ are functions of the configuration variables only, we can not construct the representation of the approximate gauge transformations as in \cref{eq:quantum-gauge-transformation}.
In the more general case, one may, for example, apply Weyl quantization techniques.
However, this may lead to other difficulties later on.
We will address these issues in a future publication.

As in the classical case, the totality of approximate lattice gauge transformations don't form a group and their generators don't form a first class system.
However, one may use the classical relation
\begin{equation}
    \diff{}{s}[\mathrlap{s=0}]\;\, \varphi_{-\sqrt{s}}^{D_n[f_n]}\circ\varphi_{-\sqrt{s}}^{D_n[g_n]}\circ\varphi_{\sqrt{s}}^{D_n[f_n]}\circ\varphi_{\sqrt{s}}^{D_n[g_n]} = \diff{}{s}[\mathrlap{s=0}]\;\, \varphi_{s}^{\poisson{D_n[f_n]}{D_n[g_n]}},
\end{equation}
and the fact that $\eta_n G_n\to 0$ to show that the corresponding relations between the quantum representations of the approximate gauge transformations will hold better and better as $n\to\infty$.

\section{The Continuum Limit}
\label{sec:The Continuum Limit}
Now that we have obtained quantized versions of our lattice theories, the logical next step is to study their continuum limit.
In the following, we will outline a method to define a continuum version of these theories.

First, we note that on every lattice Hilbert space $\mathcal H_n$, there exists a $C*$--algebra $W_n$ of Weyl elements, which is given by
\begin{equation}
    W_n = \overline{\mathrm{span}\Set{\e^{\im\hat\phi_n[f_n] + \im\hat\pi_n[g_n]} | (f_{nk})_k, (g_{nk})_k \in \mathbb R^{N_n}}} .
\end{equation}
We display the continuum Weyl algebra as the projective limit $W = \varprojlim W_n$, where the identifications
\begin{equation}
    \hat\phi_{n+1,2k}f_{n+1,2k} + \hat\phi_{n+1,2k+1}f_{n+1,2k+1} \equiv \hat\phi_{nk}(f_{n+1,2k} + f_{n+1,2k+1}),
\end{equation}
and similarly for $\hat\pi_{nk}$, are made.
We note that the Schrödinger representations on the lattice are irreducible representations of $W_n$, so every vector $\psi_n\in\mathcal H_n$ is a cyclic vector with respect to $W_n$.

In order to find a continuum limit, we need to find a sequence $(\psi_n)_n$ of normalized vectors $\psi_n\in\mathcal H_n$ such that the algebraic states
\begin{equation}
    \omega_n\left(\e^{\im\hat\phi_n[f_n] + \im\hat\pi_n[g_n]}\right) \coloneq \Braket{\psi_n, \e^{\im\hat\phi_n[f_n] + \im\hat\pi_n[g_n]} \psi_n} \label{eq:algebraic-states}
\end{equation}
converge to a state $\omega$ on $W$.
In order to check that, we need to show that the $\omega_n$ are a Cauchy sequence in the following sense:
Let $f$ be a measurable function on $\mathbb T$ and $(f_n)_n$ be a sequence of piecewise constant functions on the $n$--th lattice that converges to $f$.
Analogously, choose a sequence $g_n\to g$.
We say that the $\omega_n$ form a Cauchy sequence if for every $\epsilon>0$, there is $N>0$ such that for all $f_n\to f$, $g_n\to g$ and all numbers $n,m>N$, we have
\begin{equation}
    \abs*{\omega_n\left(\e^{\im\hat\phi_n[f_n] + \im\hat\pi_n[g_n]}\right) - \omega_m\left(\e^{\im\hat\phi_m[f_m] + \im\hat\pi_m[g_m]}\right)} < \epsilon .
\end{equation}
In that case, we define
\begin{equation}
    \omega\left(\lim_{n\to\infty} \e^{\im\hat\phi_n[f_n] + \im\hat\pi_n[g_n]}\right) \coloneq \lim_{n\to\infty} \omega_n\left(\e^{\im\hat\phi_n[f_n] + \im\hat\pi_n[g_n]}\right) .
\end{equation}
The continuum Hilbert space $\mathcal H$ and a cyclic representation of the continuum Weyl algebra $W$ with cyclic vector $\psi$ can then be reconstructed from $\omega$ using the GNS--construction.

We note that this way of obtaining a continuum theory is much simpler than the application of a renormalization group method \parencite[e.g.][]{Lang:2017beo}.
While a renormalization group flow does produce a converging sequence $(\omega_n)_n$ at the fixed point of the flow, the resulting sequence is a very complex object that is hard to obtain.
The fixed point sequence does not only converge, but it satisfies the additional property of cylindrical consistency, which means that the expectation value of a coarse Weyl element is the same on both a coarse lattice and a fine lattice.
This additional property is much stronger than mere convergence, but not strictly required for the definition of a theory in the continuum.
On the other hand, generating a convergent sequence by means of a renormalization group flow has the advantage of being quite systematic.

\subsection{Defining Convergent Sequences}
It can be difficult to choose a sequence $(\psi_n)_n$ of states that has a continuum limit.
In general, one may need to make use of the freedom to be able to choose the lattice approximations $f_n$ of the continuum test functions $f$.
Different choices may lead to different limits or may develop divergences.

We want to illustrate one particularly simple approach to obtaining such a sequence.
Let us choose an arbitrary state $\psi_{m} \in \mathcal H_{m}$.
We iteratively define a sequence of fine states $(\psi_n)_{n\geq m}$ from $\psi_{m}$ as follows:
\begin{equation}
    \psi_{n+1}((\phi_{n+1,k})_k) \coloneq 2^{-2^{n-1}} \psi_n((\phi_{nk}^+)_k) \prod_{k=1}^{N_n} \rho_{n}(\phi_{nk}^-),
\end{equation}
where
\begin{equation}
    \phi_{nk}^+ = \frac{\phi_{n+1,2k} + \phi_{n+1,2k-1}}{2}, \quad \phi_{nk}^- = \frac{\phi_{n+1,2k} - \phi_{n+1,2k-1}}{2}. \label{eq:change-of-variables}
\end{equation}
Analogously, we define $\pi_{nk}^\pm$.
The functions $\rho_n$ are arbitrary, yet to be chosen functions of one variable.
They quantify the distribution of the local fluctuations on each scale.

By expanding the canonical variables in $\hat\phi_{nk}^\pm$ and $\hat\pi_{nk}^\pm$, where $\commutator{\hat\phi_{nk}^\pm}{\hat\pi_{nk'}^\mp} = 0$, we can evaluate the algebraic states $\omega_n$ (as defined in \cref{eq:algebraic-states}) on the corresponding Weyl elements.
Using the shorthand $f_{nk}^{n+1} = f_{n+1,2k-1}+f_{n+1,2k}$, we find:
\begin{align}
    \MoveEqLeft \phi_{n+1}[f_{n+1}]
         = \sum_{k=1}^{N_{n+1}} \phi_{n+1,k}f_{n+1,k} \nonumber \\
        &= \sum_{k=1}^{N_n}\phi_{nk}^+(f_{n+1,2k-1}+f_{n+1,2k}) + \sum_{k=1}^{N_n} \phi_{nk}^-(f_{n+1,2k}-f_{n+1,2k-1}) \nonumber \\
        &= \sum_{k=1}^{N_n}\phi_{nk}^+ f_{nk}^{n+1} + \eta_{n+1} \sum_{k=1}^{N_n} \phi_{nk}^- (\Delta f)^{n+1}_{n k} \nonumber \\
        &= \phi_n^+[f_n^{n+1}] + \eta_{n+1} \,\phi_{n}^- \left[ (\Delta f)^{n+1}_{n} \right],
\end{align}
where we defined
\begin{align}
    (\Delta f)^{n+1}_{n k} \coloneq \frac{1}{\eta_{n+1}} \left( f_{n+1, 2k} - f_{n+1,2k-1}\right).
\end{align}
A similar expression follows for $\pi_{n+1}[f_{n+1}]$.
Consequently, we find the following relation for the Weyl elements:
\begin{align}
    \e^{\im\hat\phi_{n+1}[f_{n+1}] + \im\hat\pi_{n+1}[g_{n+1}]}
        = \e^{\im\hat\phi_n^+[f_n^{n+1}] + \im\hat\pi_n^+[g_n^{n+1}]} \prod_{k=1}^{N_n} \e^{\im\eta_{n+1}\left(\hat\phi_{n}^- [(\Delta f)^{n+1}_{n }] + \hat\pi_{n}^- [(\Delta g)^{n+1}_{n }]\right)},
\end{align}
We note that all factors on the right hand side commute.
This allows us to evaluate the $\omega_{n+1}$ on the Weyl elements by a simple change of variables according to \cref{eq:change-of-variables}:
\begin{align}
    \omega_{n+1}(\e^{\im\hat\phi_{n+1}[f_{n+1}] + \im\hat\pi_{n+1}[g_{n+1}]})
        = \omega_n(\e^{\im\hat\phi_n[f_n^{n+1}] + \im\hat\pi_n[g_n^{n+1}]}) \prod_{k=1}^{N_n} \zeta_{n}((\Delta f)^{n+1}_{nk}, (\Delta g)^{n+1}_{nk}) .
\end{align}
Here, we defined
\begin{equation} \label{def:zeta_n}
    \zeta_{n}((\Delta f)^{n+1}_{nk}, (\Delta g)^{n+1}_{nk}) \coloneq \Braket{\rho_n, \e^{\im \eta_{n+1} (\hat\phi_{nk}^- (\Delta f)^{n+1}_{nk} + \hat\pi_{nk}^-(\Delta g)^{n+1}_{nk})} \rho_n}.
\end{equation}
It follows that the corresponding states $\omega_n$ for any $n>m$, evaluated on elements of the Weyl algebra, read
\begin{equation}
    \omega_n\left(\e^{\im\hat\phi_n[f_n] + \im\hat\pi_n[g_n]}\right) = \omega_{m}\left(\e^{\im\hat\phi_{m}[f^n_{m}] + \im\hat\pi_{m}[g^n_{m}]}\right) \prod_{l=m}^{n-1} \prod_{k=1}^{N_l} \zeta_l ((\Delta f)^n_{m k}, (\Delta g)^n_{m k}),
\end{equation}
where we use that the components of the vectors $f^n_{m}$ and $(\Delta f)^n_m$ are respectively given by
\begin{align}
    f^n_{m k} &= \sum_{j=1}^{N_n N_{m}^{-1} } f_{n, (k-1) N_n N_{m}^{-1} +j}, \\
    (\Delta f)^n_{m k} &= \sum_{j=1}^{N_n N_{m}^{-1}} f_{n, (k-1) N_n N_{m}^{-1} +j} \cdot \sgn (2j -1- N_n N_{m}^{-1}),
\end{align}
where $k =1,\dots, N_{m}$. Similar expressions hold for $g^n_{m}$ and $(\Delta g)^n_m$.
If the $\rho_l$ are chosen appropriately, the product of the $\zeta_l$ converges and a continuum limit exists.

Possible examples that lead to convergence are given by Dirac sequences with suitable speed of convergence.
A Dirac sequence of Gaussian distributions will lead to a Gaussian state in the continuum and thus to a Fock space representation.
Dirac sequences with compact support can be used to constrain the admissible states to those with support on configurations of bounded variation.
Such states are interesting, because functions of bounded variation are differentiable almost everywhere, which can aid the convergence of finite differences $\Delta^{\eta_n} \hat\phi_{nk}$ to derivatives $\partial\hat\phi(x)$ of field operators.

\subsection{Representation of the Group of Gauge Transformations}
Now that we have a continuum Hilbert space, we would like to implement gauge transformations on it.
Unfortunately, the operators $U_{D_n[f_n]}(s) W_n U_{D_n[f_n]}(s)^\dagger$ are not necessarily Weyl elements again.
Thus, implementing the gauge transformations on the continuum Hilbert space is not as easy as taking the limit of the transformed Weyl elements in the projective limit algebra $W$.
Nevertheless, we can describe a plausible procedure:

Suppose we want to apply a gauge transformation $U_{D[f]}(s)$ on a state $\psi \in \mathcal H$ in the continuum Hilbert space.
We have to choose a sequence $(\psi_n)_n$ in the lattice Hilbert spaces $\mathcal H_n$ that converges to $\psi$ in the sense of algebraic states.
Moreover, we choose a sequence $f_n$ of piecewise constant functions with $f_n\to f$.
We can then apply the approximate gauge transformations $U_{D_n[f_n]}(s)$ to $\psi_n$.
Since every state in $\mathcal H_n$ is cyclic for $W_n$, we can expand the transformed state in every lattice Hilbert space again in terms of Weyl elements:
\begin{equation}
    U_{D_n[f_n]}(s)\psi_n = \sum_k c_{nk} \e^{\im\hat\phi_n[f_{nk}] + \im\hat\pi_n[g_{nk}]} \psi_n
\end{equation}
We can then use this sequence of Weyl elements to define the continuum gauge transformation as follows:
\begin{equation}
    U_{D[f]}(s)\psi = \lim_{n\to\infty}\sum_k c_{nk} \e^{\im\hat\phi_n[f_{nk}] + \im\hat\pi_n[g_{nk}]} \psi.
\end{equation}

\section{Conclusion} \label{sec:Conclusion}
In this paper, we presented a novel approach for taking the continuum limit of a Hamiltonian lattice regularized quantum field theory with first--class constraints.
Specifically, we considered a one--dimensional Hamiltonian theory of a scalar field that we discretized by introducing a phase space consisting of piecewise constant functions.
We assumed the first--class character of the continuous field theory to be broken by discretization in such a way that the additional anomalies scale with the lattice spacing of the regular spatial lattice.
Thereby, the evolution leading out of the constraint surface due to the second--class anomalies can be made arbitrarily small by passing to finer lattices.
We then proceeded to quantize these lattice theories using a standard Schrödinger representation.
By assuming that the constraints depend only linearly on the momenta, we obtain unitary representations of the approximate gauge transformations associated with the regularized constraints.
Since, on the level of the lattice, they are also strongly continuous, they admit generators which are the appropriate quantizations of the regularized classical constraints.
Finally, we presented a way of defining states on the algebra of continuum field operators, making use of the Cauchy sequence criterion.
Our method provides the user with considerable freedom of tuning the limit and thereby fixing desirable properties in the continuum theory.
We then provide a prescription for defining a representation of the group of gauge symmetries on the resulting continuum Hilbert space.
We expect that one can utilize the aforementioned freedom to have the continuum limit inherit the property of strong continuity of the approximate gauge transformations on the lattices.
Given the broad generality of the framework introduced in this paper, our discussion has been more of an outline than a detailed construction.
To establish concrete convergence and continuity, case--specific proofs will be necessary in more specific contexts. 
This is because the precise tuning of the continuum limit must be adapted to the particular expression for which we seek to determine the limit. 

In a next step, we intend to apply our techniques to the theory of general relativity.
In this case, the set of constraints consists of the diffeomorphism constraints and the Hamiltonian constraint.
Since the diffeomorphism constraints are linear in the momenta, we aim to utilize the methods introduced in this paper to construct a strongly continuous representation of the associated group of spatial diffeomorphisms.
This will be the topic of a future publication \parencite{gravity-continuum}, wherein missing proofs and hard estimates will be provided.
For the Hamiltonian constraint, we expect the need of more sophisticated methods as it is quadratic in the momenta and consequently, the gauge invariance breaking terms no longer tend to vanish in the limit of infinitesimal lattice spacings.
In order to produce counterterms that compensate these extra terms, renormalization group methods will become relevant.

The central theme of our proposal is to sacrifice gauge invariance on the lattice at the expense of introducing unphysical degrees of freedom at the level of the regularized theories.
Instead of fixing gauge invariance on the level of the regularized theory, our ansatz is to retain strong continuity of the approximate gauge transformations on the lattice.
On the one hand, this makes it harder to recover gauge invariance in the continuum limit.
On the other hand, we expect this to facilitate the inheritance of the strong continuity property in the continuum limit.
Hence, we find it plausible to assume that our method may, in principle, facilitate the attainment of an unprecedented achievement: a strongly continuous representation of the group of spatial diffeomorphisms on a continuum Hilbert space. Such a milestone would mark significant progress in addressing the challenging aspects of quantum gravity.

\subsection*{Acknowledgements}
We would like to express our gratitude to Bianca Dittrich for providing helpful references.
In the end, we decided to defer these references to \textcite{gravity-continuum} where they thematically fit better.
Besides, we thank Thomas Thiemann for discussions and the Insitute for Quantum Gravity at FAU Erlangen--Nürnberg for their hospitality during the final stage of this work.

Research at Perimeter Institute is supported in part by the Government of Canada through the Department of Innovation, Science and Economic Development and by the Province of Ontario through the Ministry of Colleges and Universities.

\printbibliography

\end{document}